\documentclass{jfm}

\usepackage{color}
\usepackage{graphicx}
\usepackage{natbib}
\usepackage{epstopdf}
\usepackage{amsmath}
\usepackage{esdiff} 
\usepackage{subfig}
\usepackage{float}
\usepackage{tabularx}

\ifCUPmtlplainloaded \else
  \checkfont{eurm10}
  \iffontfound
    \IfFileExists{upmath.sty}
      {\typeout{^^JFound AMS Euler Roman fonts on the system,
                   using the 'upmath' package.^^J}%
       \usepackage{upmath}}
      {\typeout{^^JFound AMS Euler Roman fonts on the system, but you
                   dont seem to have the}%
       \typeout{'upmath' package installed. JFM.cls can take advantage
                 of these fonts,^^Jif you use 'upmath' package.^^J}%
      }
  \else
  \fi
\fi

\ifCUPmtlplainloaded \else
  \checkfont{msam10}
  \iffontfound
    \IfFileExists{amssymb.sty}
      {\typeout{^^JFound AMS Symbol fonts on the system, using the
                'amssymb' package.^^J}%
       \usepackage{amssymb}%
         \let\leq=\leqslant
         
      }{}
  \fi
\fi

\ifCUPmtlplainloaded \else
  \IfFileExists{amsbsy.sty}
    {\typeout{^^JFound the 'amsbsy' package on the system, using it.^^J}%
     \usepackage{amsbsy}}
    {\providecommand\boldsymbol[1]{\mbox{\boldmath $##1$}}}
\fi

\providecommand\bnabla{\boldsymbol{\nabla}}

\newsavebox{\astrutbox}
\sbox{\astrutbox}{\rule[-5pt]{0pt}{20pt}}

\title{Frequency Response and Resonance of Elastic Hele-Shaw Cells with Application to Mechanical Filters}
\shorttitle{Frequency Response of an Elastic Hele-Shaw Cell}
\author[Arie Tulchinsky and Amir D. Gat]{Arie Tulchinsky and Amir D. Gat}
\affiliation{Faculty of Mechanical Engineering, Technion - Israel Institute of Technology, Haifa 3200003, Israel}

\pubyear{2017} \volume{999} \pagerange{999--9999}
\date{2017}
\begin{document}
\maketitle

\abstract{We study steady-state oscillations of an elastic Hele-Shaw cell excited by traveling pressure waves over its upper surface. The fluid within the cell is bounded by two asymmetric elastic sheets which are connected to a rigid surface via distributed springs and modeled by the linearized plate theory. Modal analysis yields the frequency response of the configuration as a function of three parameters: the fluidic Womersley number and two ratios of elastic stress to viscous pressure for each of the sheets. These ratios, analogous to the Capillary number, combine the effects of fluid viscosity and the sheets inertia, bending and tension. The resonance frequencies of the configuration include the resonance frequency of the upper sheet, and the resonance frequency of both sheets with a constraint of constant gap. Near the resonance frequency of the upper sheet, the fluid pressure is identical in amplitude and phase to the external excitation. For configurations where both sheets are near resonance, small changes in frequency yield significant modification of the fluidic pressure. In addition, a new resonance frequency is observed, related to the interaction between motion of fluid parallel to the elastic sheets and relative elastic displacements of the sheets. The ratio of the amplitude of the fluidic pressure to the external pressure is presented vs. frequency for several characteristic solid and fluid properties, yielding a bandpass filter behaviour. For configurations with a rigid lower surface the pressure ratio is unity at the passband, while for two elastic sheets pressure amplification or reduction occurs near the resonance frequencies. The results presented here may allow to utilize elastic Hele-Shaw configurations as protective surfaces and mechanical filters.}

\section{Introduction} 
In this work we examine the frequency response of two parallel elastic sheets containing a thin liquid film. We focus on physical regimes where viscous, elastic and inertial effects (of the fluid and solids) are of similar order of magnitude. The elastic sheets are modeled by the linear plate theory and include both bending and tension effects, and may be connected to a rigid surface via distributed springs.

Interaction of fluid viscosity with solid elasticity in Hele-Shaw cells, or similar configurations, is relevant to various research subjects. Among these are viscous peeling problems \citep{hosoi2004peeling,lister2013viscous,peng2015displacement,pihler2015displacement,carlson2016similarity}, fluid driven crack propagation \citep{roper2005buoyancy,spence1987buoyancy,lai2016elastic}, gravity currents on elastic substrates \citep{hewitt2015elastic,howell2016rivulet}, and wrinkling of lubricated sheets \citep{huang2002wrinkling,kodio2016lubricated}. At the inviscid flow limit, elastic-inertial fluid-structure-interaction have been shown to vary the solid structure resonance frequency. Initial work on the subject was pioneered by \cite{lamb1920vibrations} who examined the modified resonance frequencies of a clamped sheet in contact with a water reservoir. This work have been extended by various researchers including \cite{mclachlan1932accession}, \cite{peake1954lowest},  and \cite{de1984resonance}. Recent works on the dynamic response of elastic sheets interacting with inviscid laminar flows include \cite{crighton1991fluid}, \cite{peake2001behaviour}, and \cite{sorokin2002analysis}. 

The current work follows the approach of our previous study \citep{tulchinsky2016transient}, examining an elastic Hele-Shaw configuration as a structural element. In \cite{tulchinsky2016transient},
we studied the transient dynamics of elastic Hele-Shaw Cells due to localized excitations, assuming rigid lower surface and neglecting  elastic tension effects and inertia in both the solid and fluid. We here examine the frequency response of steady-state oscillations of two elastic plates containing a viscous film and include inertial effects of both the fluid and solid regions. Specifically, we aim to obtain extrema of fluidic pressure and solid displacements in terms of the excitation frequency and the mechanical properties of the system. 

The structure of this paper is as follows: In \S2 we define and scale the problem. In \S3 we obtain the phase and amplitude of traveling-wave solutions. In \S4 we present frequency response in terms of solid displacements and fluidic pressure for various configurations. In \S5 we discuss the results, examine new resonance frequency related to fluid motion parallel to the sheets (\S5.1), and illustrate realization of Hele-Shaw cells as mechanical filters or protective surfaces (\S5.2).

\section{Problem Formulation and Scaling}
We study steady-state oscillations of two parallel elastic sheets containing a thin liquid film. The fluid flow and solid displacements are excited by an external pressure wave with prescribed frequency and wavelength, which may be readily generalized to an arbitrary external forcing. The elastic sheets are modeled by the linear plate theory and include bending, tension, and inertial effects. 

\begin{figure}
\centering 
\includegraphics[width=1\textwidth]{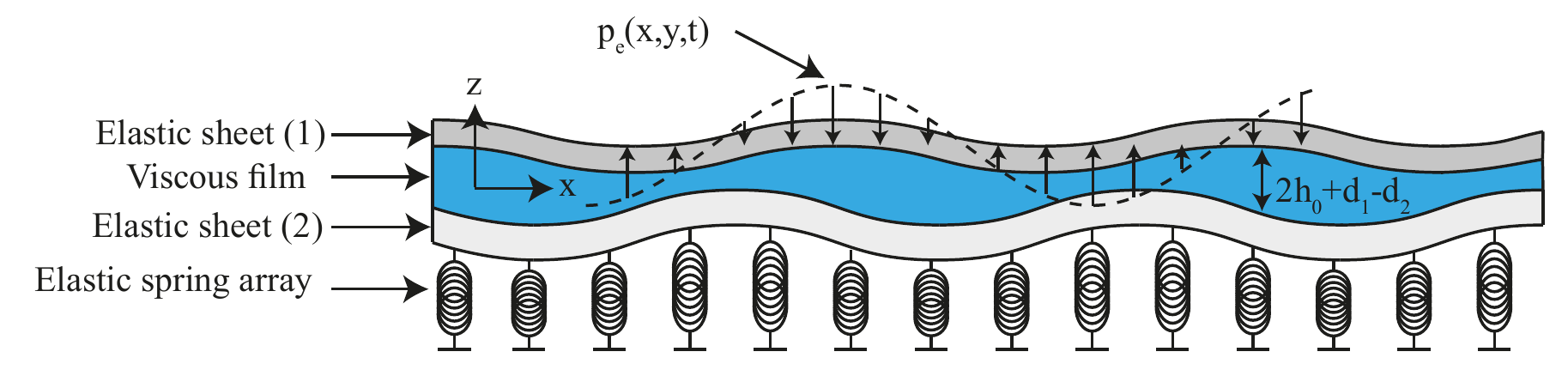}
\caption{Illustration of the configuration and the Cartesian coordinate system. Both elastic sheets are parallel at rest to the $x-y$ plane. $p_e (x,y,t)$ is the external propagating pressure wave , $2h_0$ is the gap between the sheets at rest, $d_1$ and $d_2$ are the displacements of plate 1 and 2 in the $z$-direction, respectively.}
\label{figure1}
\end{figure}

The configuration and the Cartesian coordinate system $(\boldsymbol x_\parallel, z)$ are defined in figure \ref{figure1}, where $\parallel$ subscript denotes two-dimensional vectors in the $x-y$ plane. The $x-y$ plane is parallel and of equal distance to both plates at rest. The subscripts $1$ and $2$ denote the upper and lower plates, respectively, and the gap between the plates at rest is $2 h_0$. The fluidic pressure is $p$, fluid velocity is $(\boldsymbol{u}_\parallel,w)$, fluidic density is $\rho$, fluidic viscosity is $\mu$, elastic sheet bending stiffness is $s_n$, sheet tension is $t_n$, sheet thickness is $b_n$, sheet mass-per-area is $i_n$ and sheet displacements are $(\boldsymbol d_{\parallel,n},d_n)$ (where $n=1,2$). 

The dynamics of the Newtonian, incompressible fluid is governed by the Navier-Stokes equation 
\begin{equation}
\rho\left[\frac{\partial  }{\partial t}+ (\boldsymbol u_\parallel,w) \cdot \left(\bnabla_\parallel,\frac{\partial}{\partial z}\right)\right] (\boldsymbol u_\parallel,w)=-\left(\bnabla_\parallel,\frac{\partial}{\partial z}\right) p +\mu \left(\bnabla_\parallel,\frac{\partial}{\partial z}\right)^2 (\boldsymbol u_\parallel,w), 
\label{mom_eq}
\end{equation}
and the continuity equation
\begin{equation}
\left(\bnabla_\parallel,\frac{\partial}{\partial z}\right) \cdot (\boldsymbol u_\parallel,w) =0,
\label{cont_eq}
\end{equation}
supplemented by no-slip and no-penetration at the fluid-solid interfaces
\begin{subequations}\label{kinem_bc}
\begin{align}
(\boldsymbol u_\parallel, w)&=\left(\frac{\partial \boldsymbol d_{_\parallel,1}}{\partial t} -\frac{b_1}{2} \bnabla_\parallel \frac{\partial d_1}{\partial t}, \frac{\partial d_1}{\partial t}\right),\quad z=h_0+d_1\\
(\boldsymbol u_\parallel, w)&=\left(\frac{\partial \boldsymbol d_{\parallel, 2}}{\partial t} +\frac{b_2}{2} \bnabla_\parallel \frac{\partial d_2}{\partial t}, \frac{\partial d_2}{\partial t}\right),\quad z=-h_0+d_2,
\end{align}
\end{subequations}
consistent with the Kirchhoff hypothesis of linear displacements $(d_1,d_2)$ with regard to the $z$-direction. The sheet deflections are governed by the linearized Kirchhoff-Love sheet theory \citep{howell2009applied}
 \begin{subequations}\label{kinet_bc}
\begin{align}
-s_1\nabla^4_\parallel d_1+t_1 \nabla^2_\parallel d_1 + p - p_e =i_1\frac{\partial^2 d_1}{\partial t^2},\quad z=h_0+d_1\\
-s_2\nabla^4_\parallel d_2 +t_2 \nabla^2_\parallel d_2  -s_kd_2 -p = i_2\frac{\partial^2 d_2}{\partial t^2},\quad z=-h_0+d_2,
\end{align}
\end{subequations}
where $t_1,t_2$ are considered uniform and isotropic. (The boundary conditions (\ref{kinet_bc}) may be reduced to a free surface description by setting $t_1=-\gamma$, and $s_1=i_1=0$, where $\gamma$ is surface tension.) 
The external propagating pressure wave is the real part of the function 
\begin{equation}
p_e=\hat p_e e^{i(\boldsymbol k\cdot\boldsymbol x_\parallel+ \omega t)}
\end{equation}
where $\hat p_e$ is the amplitude of the wave, $\boldsymbol k_\parallel$ is the wave vector, $k=|\sqrt{\boldsymbol k_\parallel\cdot \boldsymbol k_\parallel}|$ is the wavenumber, and $\omega$ is the angular frequency ($\hat p_e, k, \omega\in\mathbb{R}$). For the purpose of separating the flow problem from the bulk deformation of the structure, we denote hereafter the deformation of the sheets by the average $\bar d =(d_1+d_2)/2$ and relative  $d' =(d_1-d_2)/2$ deformation. We define $w'$ as 
\begin{equation}
w'=w-\frac{\partial \bar d}{\partial t},
\end{equation}
where $\partial \bar d/\partial t$ represents fluid speed due to the mean motion of both sheets and $w'$ is thus fluid speed due to the relative motion of the sheets.

Next we turn to scaling and order-of-magnitude analysis. Hereafter, asterisk superscripts denote characteristic values and Capital letters denote normalized variables. The characteristic $\boldsymbol x_\parallel$ plane fluid velocity is $u^*$, the characteristic $z$-direction fluid velocity is $w'^*$, the characteristic fluid pressure is $p^*$, the characteristic mean deformation is $\bar d^*$ and the characteristic relative deformation is $d'^*$. 

We define the following small parameters 
\begin{equation}
\varepsilon_1=h_0 k\ll1,\quad \varepsilon_2=\frac{d'^*}{h_0}\ll1
\end{equation}
corresponding to requirements of slender configuration and small relative deformation to fluid gap height. We define normalized coordinates $(\boldsymbol {X}_\parallel, Z)$  and time $T$
\begin{subequations}\label{norm_var}
\begin{equation}
(\boldsymbol {X}_\parallel, Z)= \left(\boldsymbol {x}_\parallel k, \frac{z}{h_0}\right),\quad T=t \omega,
\end{equation}
normalized mean $\bar D$ and relative $D'$ sheet deflections
\begin{equation}
\bar D=\frac{\bar d}{\bar d^*},\quad D'=\frac{d'}{d'^*}
\end{equation}
normalized fluid velocity
\begin{equation}
\boldsymbol U_\parallel=\frac{\boldsymbol u_\parallel}{u^*},\quad \bar W=\frac{\bar w}{\omega \bar d^*},\quad \bar W'=\frac{\bar w'}{\omega d'^*}
\end{equation}
and normalized fluid pressure $P$ and external pressure $P_e$
\begin{equation}
\quad P=\frac{p}{p^*},\quad P_e=\frac{p_e}{ p^*}.
\end{equation}
\end{subequations}

Substituting (\ref{norm_var}) into (\ref{mom_eq})-(\ref{cont_eq}) yields the leading order momentum equations
\begin{subequations}\label{norm_mom_eq}
\begin{equation}
\alpha^2\frac{\partial \boldsymbol U_\parallel}{\partial T}=-\bnabla P+\frac{\partial^2 \boldsymbol U_\parallel}{\partial Z^2}+O\left(\varepsilon_1^2,\varepsilon_1Re,\frac{\bar d^*}{d^{*'}}\varepsilon_1Re\right),\quad 
\label{liquid_eq}
\end{equation}
\begin{equation}\label{trans}
\alpha^2\varepsilon_1^2\left(\frac{\partial^2 D'}{\partial T^2}+\frac{\bar d^*}{d^{*'}}\frac{\partial^2 \bar D}{\partial T^2}\right)=-\frac{\partial P}{\partial Z}+O\left(\varepsilon_1^2,\varepsilon_1^3Re,\frac{\bar d^*}{d^{*'}}\varepsilon_1^3Re\right),
\end{equation}
and continuity equation
\begin{equation}
\frac{\partial W'}{\partial Z}+\bnabla\cdot\boldsymbol U_\parallel=0,
\end{equation}
\end{subequations}
where $Re=\rho_lh_0u^*/\mu$ is the Reynolds number and $\alpha^2=\rho_lh_0^2\omega/\mu$ is the Womersley number.

Order of magnitude analysis of (\ref{norm_mom_eq}) yields $d'^{*}\omega/u^*\sim\varepsilon_1$,  $p^*h_0^2k/\mu u^*\sim1$, and $\varepsilon_1Re/\alpha^2=\varepsilon_2$. Since the linear time-derivative inertial term ${\partial (\boldsymbol u_\parallel,w)}/{\partial t}$ in the LHS of (\ref{mom_eq}) scales with $\alpha^2$ while the non-linear convective terms 
$(\boldsymbol{u}_\parallel,w)\cdot (\bnabla,{\partial}/{\partial z})(\boldsymbol{u}_\parallel,w)$ scale with $\varepsilon_1Re=\alpha^2 \varepsilon_2$, we obtain the linearized Navier-Stokes momentum equations without any further assumptions. In addition, from $\varepsilon_2\ll1$ we obtain a restriction on the phase velocity of the wave $\omega/k\gg u^*$.

We substitute (\ref{norm_var}) into (\ref{kinem_bc}) and (\ref{kinet_bc}) to obtain the leading order boundary conditions
\begin{subequations}
\begin{equation}
W'(Z\sim H_0)+W'(Z\sim-H_0)=O(\varepsilon_2),
\end{equation}
\begin{equation}
W'(Z\sim H_0)-W'(Z\sim-H_0)=2\frac{\partial D'}{\partial T}+O(\varepsilon_2),
\end{equation}
\begin{equation}
\boldsymbol U_\parallel(Z\sim H_0)= O\left(b_1k\varepsilon_1,\frac{\varepsilon_1^5p^{*2}k}{2\mu\omega}\right),
\end{equation}
\begin{equation}
 \boldsymbol U_\parallel(Z\sim-H_0)= O\left(b_2k\varepsilon_1,\frac{\varepsilon_1^5p^{*2}k}{2\mu\omega}\right),
\end{equation}
\begin{equation}
\left(-s_1k^4\nabla^4_\parallel+t_1k^2 \nabla^2_\parallel-i_1\omega^2\frac{\partial^2}{\partial T^2}\right)\frac{h_0^3k^2}{\mu\omega}\left(\bar D+D'\right)+P-P_e=0
\end{equation}
and
\begin{equation}
\left(-s_2k^4\nabla^4_\parallel +t_2k^2 \nabla^2_\parallel  -s_k-i_2\omega^2\frac{\partial^2 }{\partial T^2} \right)\frac{h_0^3k^2}{\mu\omega}\left(\bar D-D'\right)-P = 0.
\end{equation}
\label{norm_bc}
\end{subequations}
In (\ref{norm_bc}c,d), $b_1k,b_2k\ll1$ are necessary conditions for applying linear plate theory and $\varepsilon_1^5p^{*2}k/2\mu\omega\ll1$ is a requirement of negligible effect of longitudinal displacements of the sheets on fluid velocity.

We focus hereafter on $\bar d^*\sim d'^*$ as well as $\alpha^2=O(1)$. Thus, the order of magnitude of the sheet displacements are $d'^*\sim h_0^3p^*k^2/\mu\omega$ and the fluidic pressure due to transverse acceleration is negligible ($\partial P/\partial Z\sim0$, see (\ref{trans})). Leading order governing equations, boundary conditions and order of magnitude for the case of dominant effect of transverse acceleration on fluidic pressure are presented in Appendix A.

\section{Phase and amplitude of traveling-wave solutions}
\label{vis_el_os}
We study steady state oscillations, and thus examine traveling-wave solutions of frequency $\omega$ and wave vector $\boldsymbol k$ equal to the external excitation pressure wave. Without loss of generality, we focus on two-dimensional configurations where the wave vector is parallel to the $x$-direction. We thus seek solutions of the form
\begin{equation}
\left(\begin{matrix} 
U\\
W\\
\bar D\\
D'\\
P\\
P_e
\end{matrix}\right)=\left(\begin{matrix} 
\hat {U}\\
\hat W\\
\hat {\bar D}\\
 {\hat D '}\\
\hat P\\
\hat P_e
\end{matrix}\right)e^{i\left(X +T\right)}.
\label{wave_var}
\end{equation}
Substituting (\ref{wave_var}) into (\ref{norm_mom_eq})-(\ref{norm_bc}), we  simplify the momentum equations (\ref{norm_mom_eq}a,b) 
\begin{subequations}
\begin{equation}
\alpha^2 i \hat U-\frac{\partial^2 \hat U}{\partial Z^2}\sim-i\hat P
\end{equation}
\begin{equation}
  \frac{\partial \hat P}{\partial Z}\sim0
\end{equation}
and continuity equation (\ref{norm_mom_eq}c) 
\begin{equation}
\frac{\partial \hat W'}{\partial Z}+i\hat U=0,
\end{equation}
\label{four_gov_eq}
\end{subequations}
as well as the boundary conditions (\ref{norm_bc}) to
\begin{subequations}
\begin{equation}
\hat W'(Z\sim H_0)\sim-\hat W'(Z\sim -H_0),
\end{equation}
\begin{equation}
\hat W'(Z\sim H_0)-\hat W'(Z\sim -H_0)\sim2i\hat D', 
\end{equation}
\begin{equation}
\hat U(Z\sim H_0)\sim\hat U(Z\sim -H_0)\sim0, 
\end{equation}
\begin{equation}
Z_1\left(\hat{\bar D}+\hat D '\right)= \hat P_e-\hat P,
\end{equation}
and
\begin{equation}
Z_2\left(\hat{\bar D}-\hat D '\right)= \hat P , 
\end{equation}
\label{four_bc_eq}
where the dimensionless parameters $Z_1$ and $Z_2$ are defined by ($n=1,2$)
\begin{equation}
Z_n=\frac{-s_nk^4-t_nk^2+i_n\omega^2-s_k(n-1)}{\mu\omega/h_0^3k^2}
\end{equation}
\end{subequations}
which may be interpreted as the ratio between the inertial and elastic stress within the solid and the traction applied by the fluid due to the viscous squeeze flow. ($Z_n$ is analogous to $1/C_a$, where $C_a$ is the Capillary number for thin films.) The limits of $Z_1\rightarrow0$ and $Z_2\rightarrow0$ correspond to resonance of elastic sheets $1$ and $2$, respectively. Negative values of $Z_1,Z_2$ are associated with dominant elastic bending and tension effects whereas positive values are associated with dominant inertial effects.

We initially solve (\ref{four_gov_eq}\textit{b}) together with (\ref{four_bc_eq}\textit{c,d}) to obtain the longitudinal fluid velocity
\begin{equation}
\hat U=\left(\frac{\cosh(\sqrt{\alpha^2i}Z)}{\cosh(\sqrt{\alpha^2i}H_0)}-1\right)\frac{\hat P}{\alpha^2}.
\label{uhat_sol}
\end{equation}
We substitute (\ref{uhat_sol}) into (\ref{four_gov_eq}\textit{c}), integrate with respect to $Z$, and apply (\ref{four_bc_eq}\textit{a}) to obtain the transverse liquid velocity 
\begin{equation}
\hat W'=i\left(Z-\frac{\sinh(\sqrt{\alpha^2i}Z)}{\sqrt{\alpha^2i}\cosh(\sqrt{\alpha^2i}H_0)}\right)\frac{\hat P}{\alpha^2}.
\label{wtag_sol}
\end{equation}
Substituting into (\ref{four_bc_eq}\textit{b}) we obtain the relative sheet deflection
\begin{equation}
\hat D'=\left(H_0-\frac{\tanh(\sqrt{\alpha^2i}H_0)}{\sqrt{\alpha^2i}}\right)\frac{\hat P}{\alpha^2}.
\label{dtag_sol}
\end{equation}
We substitute (\ref{dtag_sol}) into (\ref{four_bc_eq}\textit{f}) and obtain the mean sheet deformation
\begin{equation}
\hat {\bar D}=\left(\frac{1}{\alpha^2}\left(H_0-\frac{\tanh(\sqrt{\alpha^2i}H_0)}{\sqrt{\alpha^2i}}\right)+\frac{1}{Z_2}\right)\hat P.
\label{dbar_sol}
\end{equation}

Finally, we subtract (\ref{four_bc_eq}\textit{e,f}) and substitute (\ref{dtag_sol}) to obtain the liquid pressure 
\begin{equation}
\hat P=\left(\frac{2Z_1}{\alpha^2}\left(H_0-\frac{\tanh(\sqrt{\alpha^2i}H_0)}{\sqrt{\alpha^2i}}\right)+1+\frac{Z_1}{Z_2}\right)^{-1}\hat P_e.
\label{phat_sol}
\end{equation}
We thus obtain $\hat U, \hat W', \hat D', \hat {\bar{D}}$ and $\hat P$, representing the steady-state dynamics by a complex amplitude which is a function of the Womersley number $\alpha^2$ and the sheets impedance $Z_1$ and $Z_2$. 
Equation (\ref{dbar_sol}) yields that the order of magnitude relation $d'^*/{\bar d}^*\sim1$ defined after (\ref{norm_bc}f) becomes invalid at the limit $Z_2\rightarrow0$. This occurs since fluidic effects are negligible at this limit, which negates the assumptions of the scaling scheme. Solutions (\ref{uhat_sol})-(\ref{dbar_sol}) in their dimensional form may be found in appendix \ref{app_dim_res}, as well as the three-dimensional response dynamics for an arbitrary external pressure field.

For the limit of negligible fluidic inertial effects $\alpha^2\rightarrow0$
(\ref{uhat_sol})-(\ref{phat_sol}) may be further simplified. In this limit, the liquid pressure is 
\begin{equation}
\hat P\sim\left(1+i\frac{2Z_1H_0^3}{3}+\frac{Z_1}{Z_2}\right)^{-1} \hat P_e,
\end{equation}
the mean and relative deformations of the sheets are 
\begin{equation}
\hat D'\sim \frac{iH_0^3\hat P}{3}, \quad \hat {\bar D}\sim\left(\frac{i H_0^3}{3}+\frac{1}{Z_2}\right)\hat P
\end{equation}
and the fluid longitudinal and transverse speeds are
\begin{equation}
\hat U\sim \frac{i(Z^2-H_0^2)\hat P}{2},\quad \hat W'\sim\frac{(Z^3-3 Z H_0^2)\hat P}{6}. 
\end{equation}

\begin{figure}
\centering 
\includegraphics[width=\textwidth]{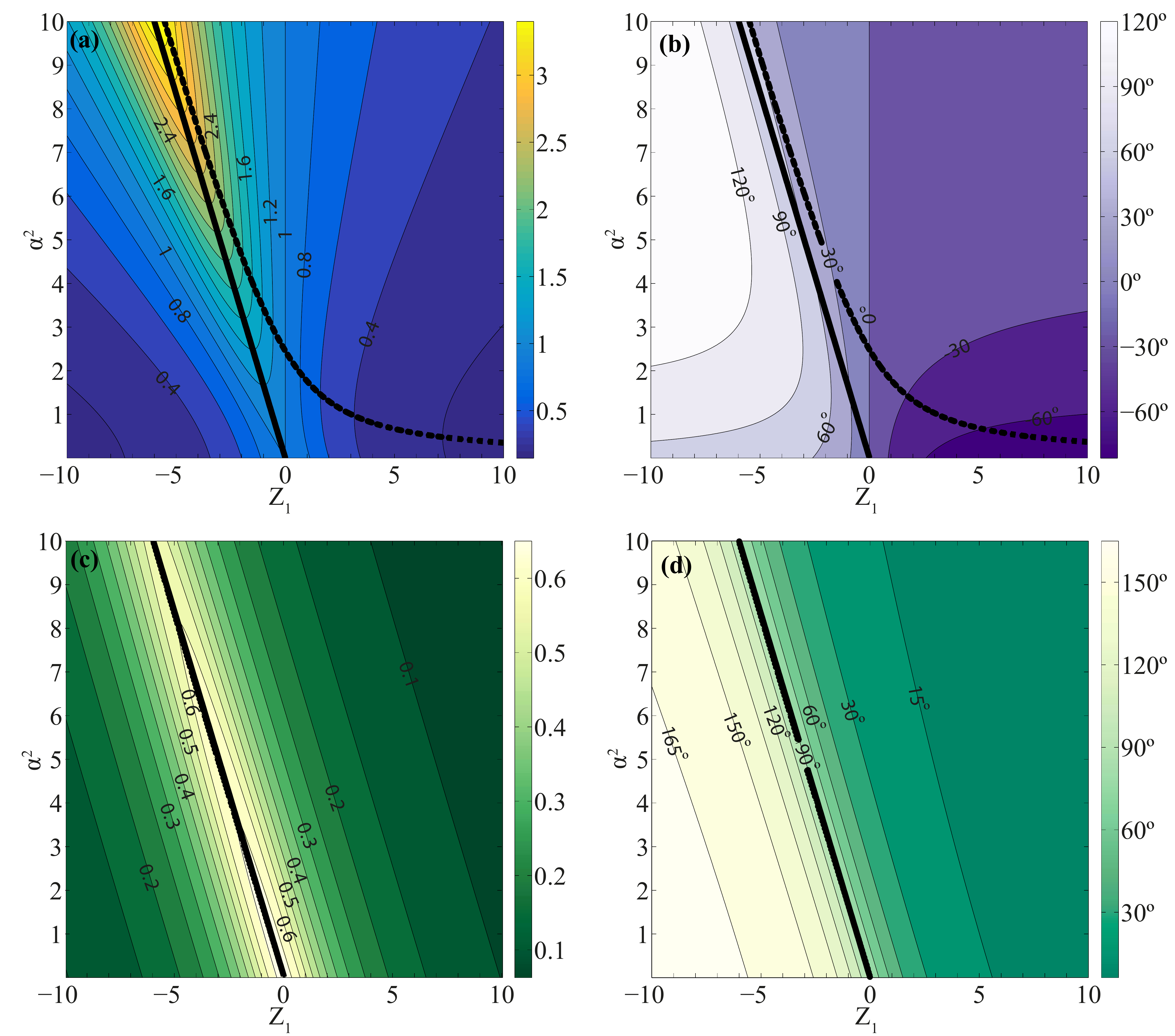}
\caption{Dynamics of a configuration with a lower rigid surface, i.e. $Z_2\rightarrow\pm\infty$. (a) is the pressure ratio magnitude $|P/P_e|$, (b) is the phase between the liquid pressure and the external pressure $\angle P-\angle P_e$, (c) is the normalized magnitude of deformation $|D_1/P_e|$, and (d) is the phase of between the upper sheet deformation and the external pressure $\angle D_1-\angle P_e$. The solid curve, defined by $\alpha^2\sim-5Z_1/3H_0$, indicates maximal magnitude of deformation and liquid pressure for constant $\alpha^2$. The dashed curve indicates maximal magnitude of liquid pressure for constant $Z_1$ (this curve coincides with the solid curve in panels c,d).}
\label{figure2}
\end{figure}

\section{Maps and extrema lines of fluid pressure and elastic displacements}
\begin{figure}
\centering 
\includegraphics[width=\textwidth]{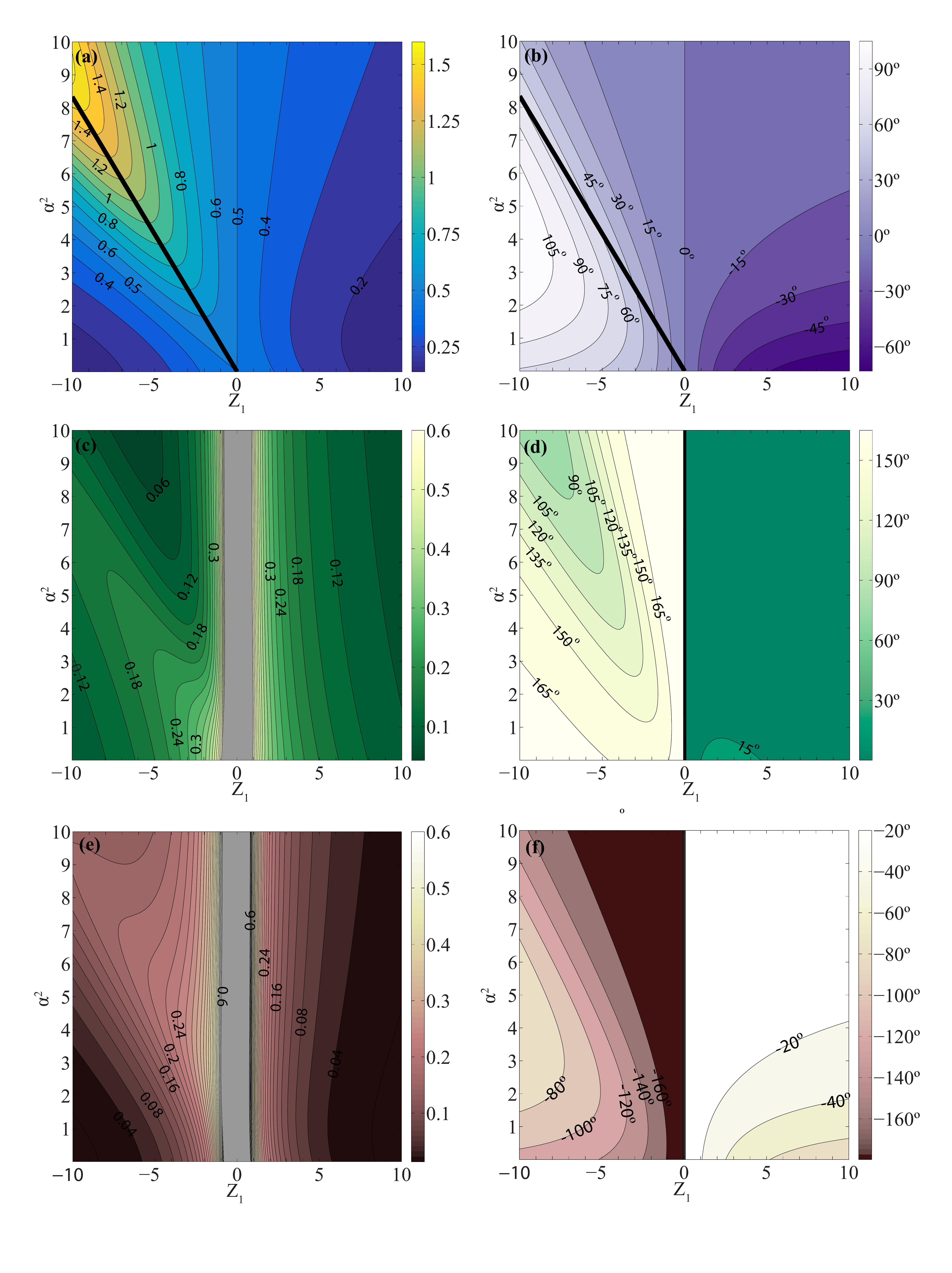}
\caption{Dynamics of a configuration consisting of two  sheets with identical impedance, $Z_1=Z_2$. Panel (a) shows the magnitude of pressure ratio  $|P/P_e|$, panel (b) shows the phase between the liquid pressure to external pressure  $\angle P-\angle P_e$, panel (c) shows the magnitude of deformation of the upper sheet $|D_1/P_e|$, panel (d) shows the phase of the deformation field with respect to the external pressure $\angle D_1-\angle P_e$, panel (e) shows the magnitude of lower plate deformation $|D_2/P_e|$, and panel (f) shows the phase of lower plate deformation to external pressure $\angle D_2-\angle P_e$. Gray colored area indicates that the value of the variable exceeds the maximal value of the color bar and is singular at $Z_1\sim0$. The smooth line denotes values of $Z_1$ yielding extrema for predefined $\alpha^2$.}
\label{symfig}
\end{figure}

We here present and discuss the frequency response relations obtained in \S3 and map (in figures \ref{figure2}-\ref{dfig}) the amplitude and phase of liquid pressure and sheet deflection for various configurations. In all figures smooth lines represent values of $Z_1$ yielding extrema of the amplitude of the examined parameter for set values of $(\alpha^2,Z_2)$. Similarly, dotted lines represent values of $\alpha^2$ yielding extrema points for set values of $(Z_1,Z_2)$.

Extrema of the fluidic pressure are readily obtained from (\ref{uhat_sol})-(\ref{phat_sol}), where for simplicity we seek extrema of $|P(\alpha^2,Z_1,Z_2)/P_e|$ by examining $|P_e/P|^2$, defined by 
\begin{equation}
\bigg|\frac{P_e}{P}\bigg|^2=\left(2Z_1F_1+1+\frac{Z_1}{Z_2}\right)^2+\left(2Z_1 F_2\right)^2,
\label{pe_opt}
\end{equation}
where 
\begin{equation}
F_1(\alpha)=\frac{1}{\alpha^2}-\frac{2}{\left(2\alpha^2\right)^{\frac{3}{2}}}\left(\frac{\sinh\left(\sqrt{\alpha^2/2}\right)+\sin\left(\sqrt{\alpha^2/2}\right)}{\cosh\left(\sqrt{\alpha^2/2}\right)+\cos\left(\sqrt{\alpha^2/2}\right)}\right).
\end{equation}
and
\begin{equation}
F_2(\alpha)=\frac{2}{\left(2\alpha^2\right)^{\frac{3}{2}}}\left(\frac{\sinh\left(\sqrt{\alpha^2/2}\right)-\sin\left(\sqrt{\alpha^2/2}\right)}{\cosh\left(\sqrt{\alpha^2/2}\right)+\cos\left(\sqrt{\alpha^2/2}\right)}\right).
\end{equation}
By differentiation of equation (\ref{pe_opt}) we obtain values of $Z_1$ yielding extrema of (\ref{pe_opt}) for predefined $(Z_2,\alpha^2)$  
\begin{subequations}
\begin{equation}\label{Z_2_alpha_2_const}
Z_1=-\left(2F_1+\frac{1}{Z_2}\right)\times \left[\left(2 F_1+\frac{1}{Z_2}\right)^2+4 F_2^2\right]^{-1}.
\end{equation}
Similarly, values of $Z_2$ yielding extrema of (\ref{pe_opt}) for predefined $(Z_1,\alpha^2)$ are
\begin{equation}
{Z_2}=-\frac{Z_1}{2Z_1 F_1+1}. 
\label{z1_max_diff_z2}
\end{equation}
Combining (\ref{z1_max_diff_z2}) into (\ref{Z_2_alpha_2_const}) yields the extrema $Z_1=Z_2=0$ for a predefined value $\alpha^2$, in which $|P(\alpha^2,Z_1,Z_2)/P_e|$ is singular and is determined by the limit of the ratio $Z_1/Z_2$. 

Differentiating (\ref{pe_opt}) with respect to $\alpha^2$, an implicit relation for extrema in terms of $\alpha^2$ for predefined values of both $Z_1$ and $Z_2$ is given by
\begin{equation}
Z_1\bigg[Z_1\frac{\partial }{\partial \alpha^2}\left(F_1^2+F_2^2\right)+\frac{\partial F_1}{\partial \alpha^2}\left(1+\frac{Z_1}{Z_2}\right)\bigg]=0.
\label{opt_max}
\end{equation}
\end{subequations}
Expressions for the extrema of the amplitude of the average $\hat {\bar D}(Z_1,Z_2,\alpha^2)$ and relative $\hat D'(Z_1,Z_2,\alpha^2)$ solid displacements are presented in Appendix C.



\subsection{Rigid lower surface, $Z_2\rightarrow\pm\infty$}
Figure (\ref{figure2}) presents (\ref{dtag_sol}) and (\ref{phat_sol}) for the limits of $Z_2\rightarrow\pm\infty$, corresponding to a fixed lower surface $D_2\sim0$. Panel (a) presents the magnitude ratio of liquid pressure to external pressure $|P/P_e|$, panel (b) shows the relative phase of liquid pressure  $\angle P-\angle P_e$, panel (c) shows the magnitude of deformation normalized by the external pressure $|D_1/P_e|$, and panel (d) shows the relative phase of deformation $\angle D_1-\angle P_e$. 

Panels (a) and (c) in figure (\ref{figure2}) present maxima of the liquid pressure as well as sheet deformation along a smooth black line defined by
\begin{equation}
\label{linear_max}
\alpha^2\sim-\frac{5Z_1}{3H_0}
\end{equation}
(or in dimensional form $\omega^2\sim 5h_0^3(s_1k^6+t_1k^4)/(3\rho_lh_0^2+5i_1h_0^3k^2)$),
representing linearization of the extrema condition (\ref{Z_2_alpha_2_const}) with regard to $\alpha^2$ around $\alpha^2=0$ for $|Z_2|\rightarrow\infty$.

Fluid inertia thus reduces the resonance frequency of the configuration, which emanates from increasing the mass accelerated during oscillations. 
On the line $\alpha^2\sim-5Z_1/3H_0$, the fluidic pressure decreases with $\alpha^2$,  the amplitude of deformation increases with $\alpha^2$, and the sheet velocity is synchronized with the external pressure $\angle W_1'-\angle P_e=\pi$.
Thus for a given deformation amplitude, external oscillating pressure waves apply the maximal external work on the sheet, and thus maximal dissipation, at the line (\ref{linear_max}). The sheet deformation reaches a global maxima for $\alpha^2=0$ and $Z_1=0$, corresponding to the resonance frequency of the upper sheet and negligible fluid inertia parallel to the sheets. At the resonance frequency of the upper plate (on the line $Z_1=0$), the fluidic Womersley number $\alpha^2=\rho_lh_0^2\omega/\mu$ does not affect the liquid pressure which is equal to the external pressure in both magnitude and phase. In addition, the liquid pressure amplitude is equal to the external pressure on the line $\alpha^2\sim-5H_0Z_1/6$, on which however $\angle W_1'-\angle P_e\neq \pi$.

\subsection{Equal impedance of the lower and upper sheets, $Z_1=Z_2$}
Figure \ref{symfig} shows dynamics for configurations where $Z_1=Z_2$ (in dimensional terms $-s_1k^4-t_1k^2+i_1\omega^2=
-s_2k^4-t_2k^2+i_2\omega^2-s_k$). Panel (a) presents $|P|/|P_e|$, panel (b) shows $\angle P-\angle P_e$, panel (c) shows $|D_1|/|P_e|$, panel (d) shows $\angle D_1-\angle P_e$, panel (e) shows $|D_2|/|P_e|$, and panel (f) shows $\angle D_2-\angle P_e$.

For predefined values of $\alpha^2$ and ratio of sheet impedance $Z_1/Z_2\equiv R_{12}$, values of $Z_1$ yielding extrema of fluidic pressure are defined by
\begin{equation}
Z_1=-\frac{(1+R_{12})F_1(\alpha)}{2F_1^2(\alpha)+2F_2^2(\alpha)},\label{aaaa}
\end{equation}
which can be approximated by $Z_1\approx-3(1+R_{12})\alpha^2/5H_0$ (presented by smooth lines in figure \ref{symfig}). The limit of $(Z_1,Z_2)\rightarrow(0,0)$ yields various values of pressure and displacements depending on the ratio of $Z_1/Z_2$. For $\lim_{(Z_1,Z_2)\rightarrow(0,0)}Z_1/Z_2=-1$, resonance dynamics are accompanied with significant increase in the magnitude of the fluidic pressure. In contrast, for $\lim_{(Z_1,Z_2)\rightarrow(0,0)}Z_1/Z_2=1$ the fluidic pressure equals half of the external pressure excitation. 

Panel (a) presents the extrema line (\ref{aaaa}) which is accompanied by a maxima of the relative displacements $|D'|/| P_e|$. This maxima line is similar to the modified resonance of the upper plate presented in figure 2a. Panels (c) and (e) present the solid resonances along $Z_1=Z_2=Z_1+Z_2=0$ and an additional fluidic maxima of the upper plate near the line (\ref{aaaa}).
Panels (d) and (f) yield a sharp phase difference of $D_1$ and $D_2$ between $Z_1>0$ and $Z_1<0$. Both $D_1$ and $D_2$ are near anti-phase for small negative $Z_1$ and in-phase for small positive $Z_1$. In contrast, the phase of the fluidic pressure (panel b) presents gradual change near resonance. Panels (a), (c) and (e) show that the resonance of the solid deflection for $Z_1=Z_2\rightarrow 0$, yields similar oscillations of the upper and lower sheets. These oscillations are accompanied by synchronous liquid pressure with half the magnitude of the external pressure ($|P|/|P_e|=1/2$ and $\angle P-\angle P_e=0$). Thus, both sheets are applied with identical excitations at resonance.

\begin{figure}
\centering 
\includegraphics[width=\textwidth]{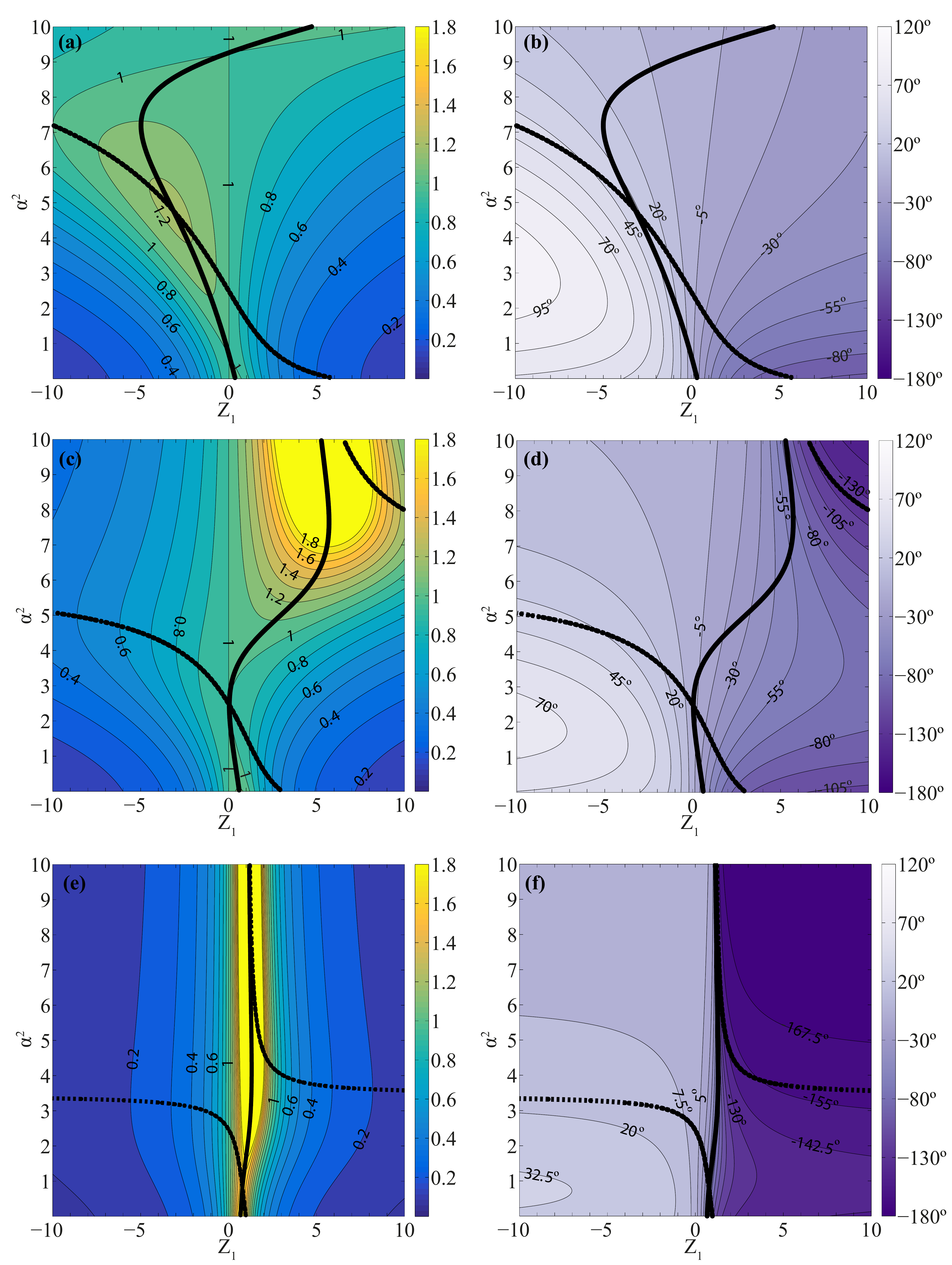}
\caption{Magnitude ratio and phase  between the liquid pressure to external pressure as a function of Womersley number $\alpha^2=\rho_lh_0^2\omega/\mu$, and upper sheet impedance parameters $Z_1=(-s_1k^4-t_1k^2+i_1\omega^2)h_0^3k^2/\mu\omega$. The left column (a,c,e) shows the magnitude ratio $|\hat P/ \hat P_e|$. The right column (b,d,f) shows the phase ratio $\angle \hat P-\angle P_e$. Panels (a,b) present $Z_2=-6$; panels (c,d) present $Z_2=-3$, and panels (e,f) present $Z_2=-1$. 
The smooth (\ref{Z_2_alpha_2_const}) and dotted lines (\ref{z1_max_diff_z2}) present extrema of deflection with regard to $Z_1$ (for set values of $\alpha^2,Z_2$) and with regard to $\alpha^2$ (for set values of $Z_1,Z_2$), respectively.}
\label{pfig}
\end{figure}

\begin{figure}
\centering 
\includegraphics[width=\textwidth]{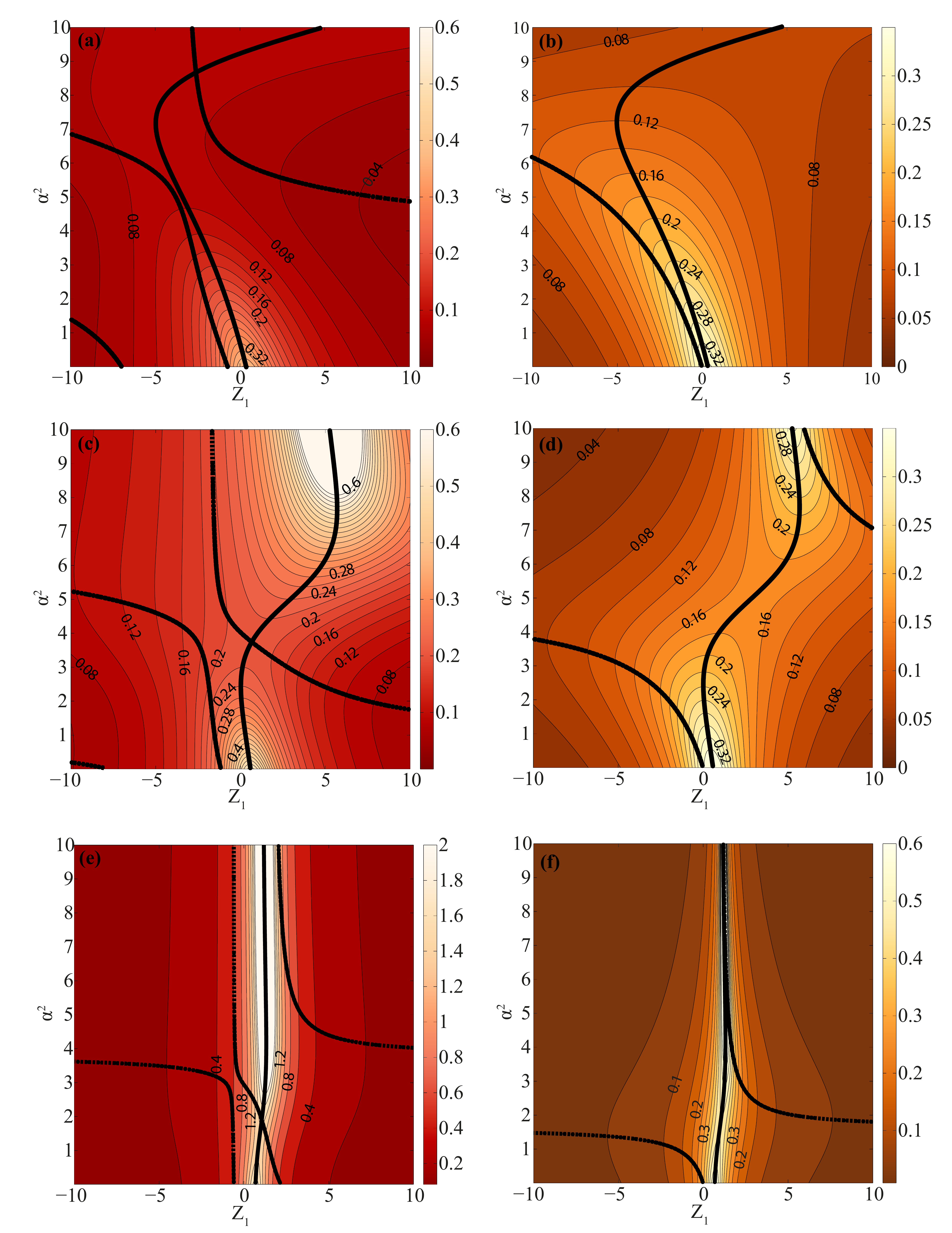}[h]
\caption{Scaled average deformation magnitude $|\bar D|/|P_e|$  (a,c,e), and scaled relative deformation $|D'/|P_e|$ (b,d,f), as a function of Womersley number $\alpha^2$, and the upper sheet impedance $Z_1$. Panels (a,b) present $Z_2=-6$; panels (c,d) present $Z_2=-3$, and panels (e,f) present $Z_2=-1$. 
The smooth and dotted lines  present extrema of deflection with regard to $Z_1$ (for set values of $\alpha^2,Z_2$) and with regard to $\alpha^2$ (for set values of $Z_1,Z_2$), respectively (see Appendix C).}
\label{dfig}
\end{figure}

\subsection{Asymmetric elastic sheets}
In figures (\ref{pfig}) and (\ref{dfig}) we examine the effect of modifying the properties of the lower surface on the frequency response and extrema for configurations where $Z_2=-6,-3,-1$ for the range $-10\leq Z_1 \leq 10$ and $0 \leq \alpha^2 \leq 10$. 

Figure \ref{pfig} presents scaled fluid pressure amplitude $|P/P_e|$ (left column)  and relative phase $\angle P-\angle P_e$ (right column). Panels (a,b) correspond to $Z_2=-6$, panels (c,d) correspond to $Z_2=-3$, and panels (e,f) correspond to $Z_2=-1$. The smooth lines (\ref{Z_2_alpha_2_const}) represent values of $Z_1$ yielding extrema of the liquid pressure amplitude for set values of $(\alpha^2,Z_2)$. The dotted lines (\ref{opt_max}) present, similarly, values of $\alpha^2$ for set values of $(Z_1,Z_2)$. 

For $Z_2=-6$, panel (a) presents maxima and saddle points along (\ref{opt_max}) where the maxima associated with $\alpha^2=0$ is on $Z_1=-(1/Z_2+4/9)^{-1}$, not on $Z_1=0$. In this case the fluid inertia decreases the resonance frequency, similarly to the cases presented in figures 2 and 3. However, as $Z_2$ increases additional maxima emerge (see panel c) and coalesce (see panel e) to a single line near $Z_1\approx-(1/Z_2+4/9)^{-1}$  in which the magnitude of the derivative of fluid pressure amplitude and phase with regard to $Z_1$ sharply increases. This new extremum emanates from a combined fluid-elastic interaction increasing the fluidic pressure (see equation (\ref{phat_sol}) and not a modified solid resonance. As $Z_2$ becomes smaller, configurations with positive values of $Z_1$ become increasingly synchronized with the external pressure and the liquid pressure, whereas configurations with negative $Z_1$ approach the inverse phase of $-180^0$. The fluidic pressure amplitude is equal to the external amplitude on two curves, one of which is the upper sheet resonance frequencies line $Z_1=0$.

Figure \ref{dfig} focuses on elastic deflection and presents the magnitude of average deformation  $|\bar D|/|P_e|$ (left column), and the relative deformation $|D'|/|P_e|$ (right column) as a function of $\alpha^2$ and $Z_1$ for identical parameter range as in figure \ref{pfig}. The smooth and dotted lines represent extrema of deflection with regard to $Z_1$ (for set values of $\alpha^2,Z_2$) and with regard to $\alpha^2$ (for set values of $Z_1,Z_2$), respectively, see Appendix C. The patterns presented for both the scaled average deflection $|\bar D|/|P_e|$ and scaled relative deflection $|D'|/|P_e|$ closely follow that of the fluidic pressure. The average deflection $|\bar D|/|P_e|$ does not involve viscous flow and is thus undamped, in contrast with $|D'|/|P_e|$. Hence, while far from resonance the magnitude of $|\bar D|/|P_e|$ and  $|D'|/|P_e|$ are similar, near resonance frequencies (e.g. near the line $Z_1\approx-(1/Z_2+4/9)^{-1}$ for $Z_2=-1$), the ratio $|\bar D|/|D'|$ increases indefinitely and 
eventually invalidates the model assumption of $O(|\bar D|)\sim O(|D'|)$.

\section{Discussion}
We here focused on configurations where the average and relative displacements are of similar order of magnitudes $O(\hat D')\sim O(\bar{\hat D})$. In the limit of $\hat D'\ll\bar{\hat D}$ (presented in Appendix A), although liquid pressure is created by elastic displacements, it is not significant in determining the  displacement dynamics. The opposite limit of  $\hat D'\gg\bar{\hat D}$ is not physically possible, as evident directly from the ratio  
\begin{equation}
\bigg| \frac{\bar{\hat D}}{\hat D'}\bigg|^2=1+\frac{2F_1}{(F_1^2+F_2^2)Z_2^2}+\frac{F_1^2+F_2^2}{(F_1^2+F_2^2)^2Z_2^4},
\label{mag_dbar_to_dtag}
\end{equation}
where all the terms of the RHS of Eq. (\ref{mag_dbar_to_dtag}) are positive.

In \S4 the response of an elastic Hele-Shaw cell was presented in terms of the parameters $(\alpha,Z_1,Z_2)$, which combine effects of elasticity, viscosity, fluid and solid inertia, as well as the frequency and wavelength of the excitation. While figures 2-5 describe a wide range of parameters, the effect of changing excitation frequency or wavelength of a specific configuration requires following curved lines (e.g. on $Z_n=(-s_nk^4-t_nk^2+i_n\omega^2-s_k(n-1))/(\mu\omega/h_0^3k^2)$, $n=1,2$ in figures 2 and 3) or interpolation between panels (as in figures 4 and 5). Clarity might thus benefit from explicit presentation of the amplitude of the fluidic pressure and elastic displacement vs. the excitation frequency for specific and constant physical parameters. In addition, the effect of the fluid will be emphasized here  by comparison with relevant fully elastic configurations. 


\begin{figure} 
\centering 
\includegraphics[width=\textwidth]{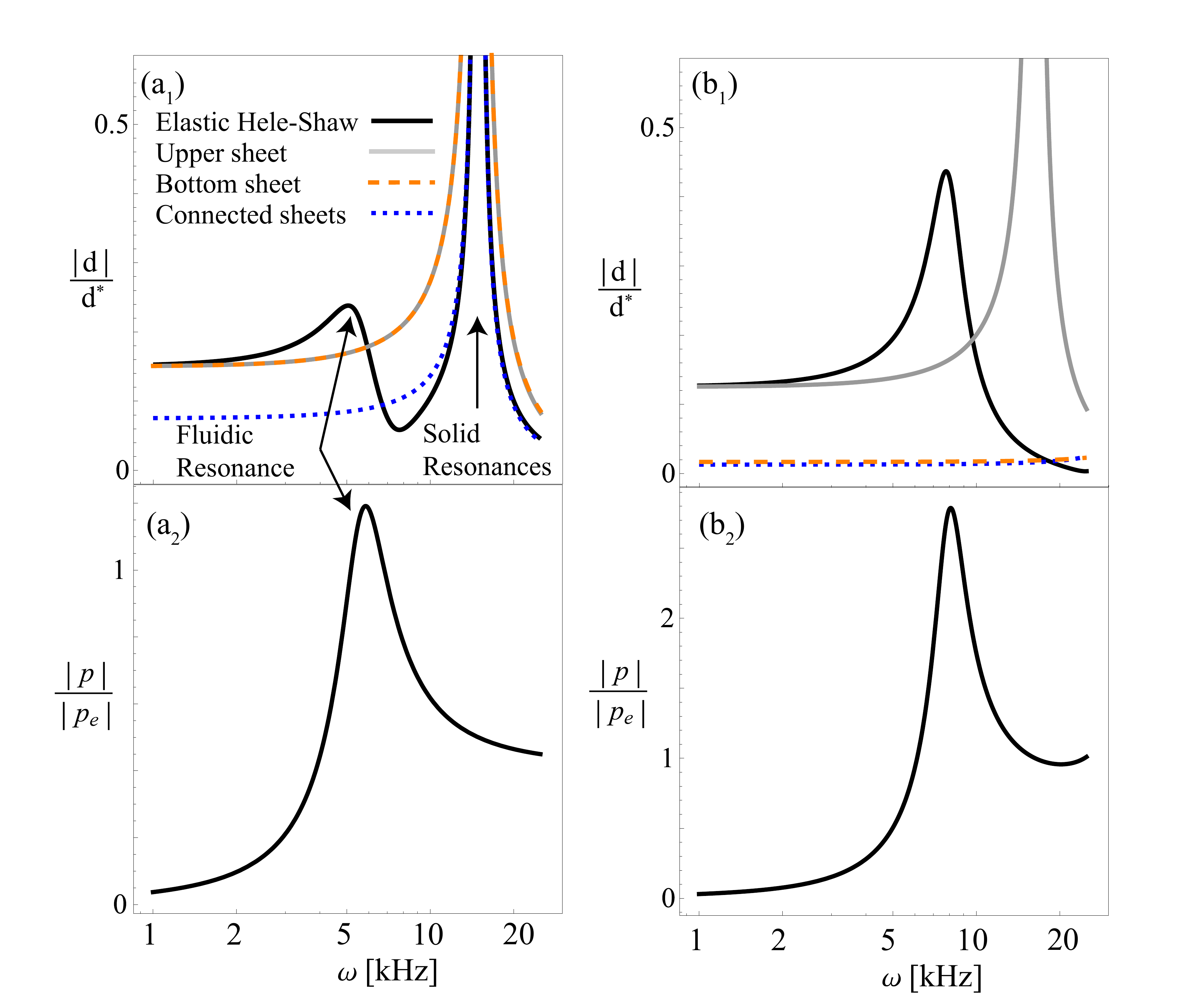}
\caption{Illustration of the additional resonance frequency emerging due to parallel motion of fluid. Panels ($a_1,a_2$) examines equal sheet impedance $Z_1=Z_2$ (defined by $\rho_1=\rho_2=10^3Kg/m^3$, $s_1=s_2=0.054 Pa/m^3$, $i_1=i_2=1Kg/m^2$, $\mu=1Pa\cdot s$, $\rho_l=10^3Kg/m^3$, $h_0=10^{-3}$, $2b_1=2b_2=0.001m$). Panels ($b_1,b_2$) examine a rigid lower sheet, $Z_2 \gg Z_1$, (defined by $\rho_1=\rho_2=10^{3}Kg/m^3$, $s_1=0.0065Pa/m^3$, $s_2=0.057Pa/m^3$, $i_1=i_2=1Kg/m^2$, $\mu=1Pa\cdot s$, $\rho_l=10^3Kg/m^3$, $h_0=10^{-3}$, $2b_1=2b_2=10^{-3}m$). 
Normalized displacement of the upper sheet ($a_1$, $b_1$) and fluidic pressure ($a_2$, $b_2$) for an elastic Hele-Shaw are marked by smooth black lines. For comparison, the isolated solid response of the upper sheet (grey lines), the bottom sheet (dashed orange lines) and the two sheets with a constraint of constant gap (blue dashed) are presented.}
\label{resfig}
\end{figure}

\subsection{The emergence of an additional fluid-elastic resonance frequency}
In the limit of negligible fluidic effects 
the dynamics of the upper surface will approach the dynamics of an isolated sheet. Alternatively, in the limit of a highly viscous fluid 
the configuration will be similar to two elastic sheets with a constraint of constant gap (see Appendix D). Thus, we expect multiple elastic resonance frequencies defined by $Z_1=0$, $Z_2=0$ and $Z_1+Z_2=0$ to appear. However, from figures 2-5 an additional response frequency is evident, which involves the interaction between motion of fluid parallel to the elastic sheets and elastic displacements and external actuation perpendicular to the sheets.

This fluid-elastic resonance is presented in figure \ref{resfig} for two illustrative configurations. Panel (a) examines the normalized displacement of the upper sheet ($a_1$) and fluid pressure amplitude ($a_2$) of identical sheets $Z_1=Z_2$ (defined by $\rho_1=\rho_2=10^3Kg/m^3$, $s_1=s_2=0.054 Pa/m^3$, $i_1=i_2=1Kg/m^2$, $\mu=1Pa\cdot s$, $\rho_l=10^3Kg/m^3$, $h_0=10^{-3}$, $2b_1=2b_2=0.001m$). Panel (b) similarly presents upper sheet normalized displacement ($b_1$) and fluid pressure amplitude ($b_2$) for configuration where $Z_2 \gg Z_1$, (defined by $\rho_1=\rho_2=10^{3}Kg/m^3$, $s_1=0.0065Pa/m^3$, $s_2=0.057Pa/m^3$, $i_1=i_2=1Kg/m^2$, $\mu=1Pa\cdot s$, $\rho_l=10^3Kg/m^3$, $h_0=10^{-3}$, $2b_1=2b_2=10^{-3}m$). Normalized deformation magnitude of the upper plate of elastic Hele-Shaw cell is marked by black smooth lines. Grey, dashed-orange and dotted-blue lines mark relevant fully elastic configurations of the upper sheet, lower sheet and connected elastic sheets, respectively. All deformation are scaled by $d^*=3\cdot10^{-6}m$.

For both configurations presented in panels ($a$) and ($b$), at lower frequencies corresponding to $|Z_1|\gg 1$, the amplitude of fluid pressure decreases with $\omega$ and displacement of the top sheet is identical to that of an isolated elastic sheet. For the opposite limit of large $\omega$, corresponding to $|Z_1|,|Z_2|\ll 1$, the elastic Hele-Shaw cell oscillates as two elastic sheets with a constraint of constant gap (see Appendix D). For panel (a), in accordance with the results presented in figure 3, the identical properties of the bottom and top sheets yield a single solid resonance frequency $\omega\approx14.9KHz$ accompanied by $|p/p_e|=0.5$. In addition to this elastic resonance frequency, a clear additional extrema frequency is evident at $\omega\approx5.05KHz$, near the extrema of the fluidic pressure at $\omega\approx5.05KHz$. A similar deformation extrema is presented in panel (b) for $\omega\approx7.77KHz$, near the pressure extrema of $\omega\approx8.09KHz$, in which it is the dominant resonance frequency due to the rigid lower elastic sheet eliminating other solid resonances. 

\subsection{Elastic Hele-Shaw Cell as a Mechanical Filter}
\begin{figure}
\centering 
\includegraphics[width=0.7\textwidth]{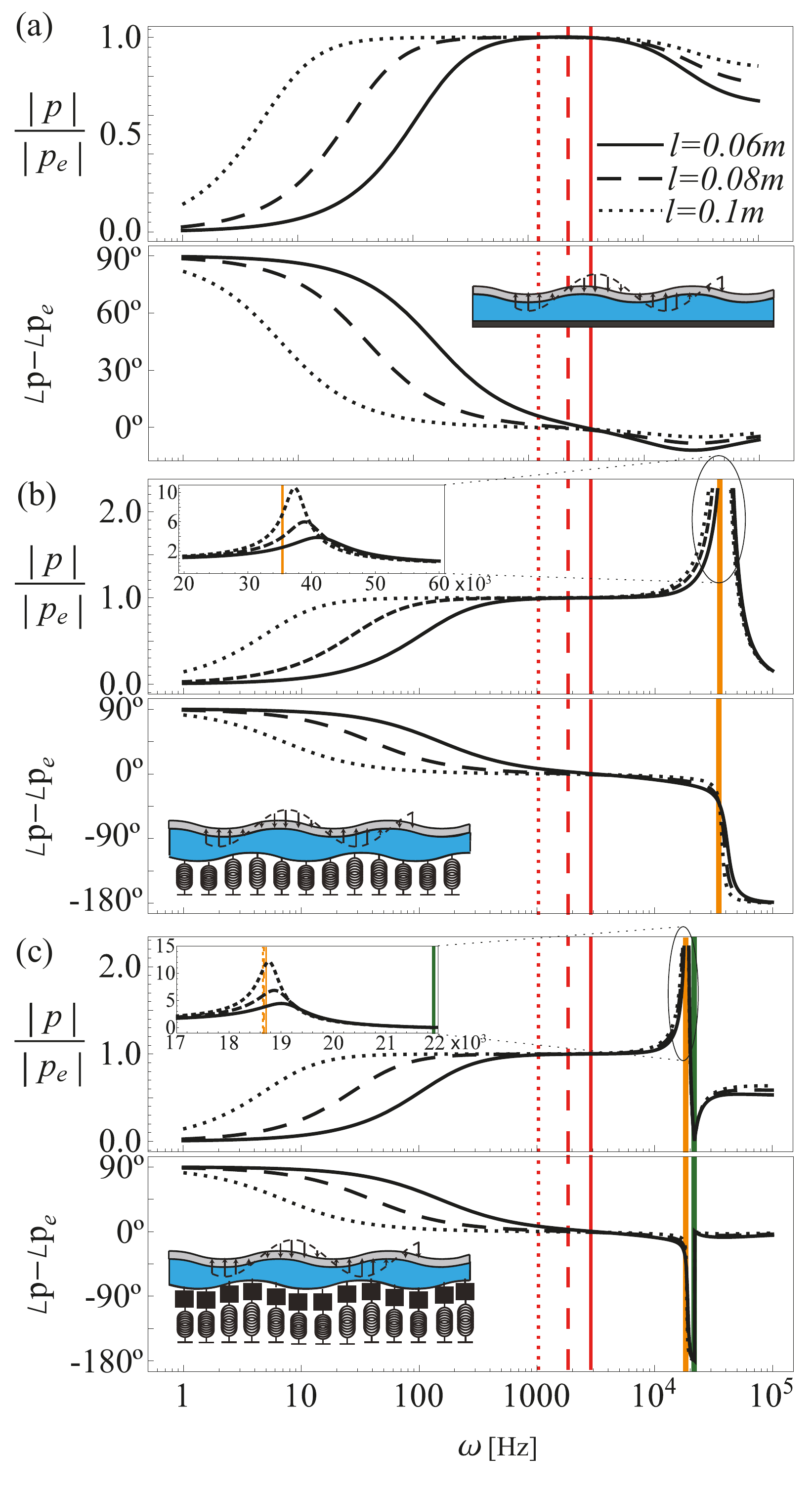}
\caption{Frequency response of elastic Hele-Shaw cells for three physical configurations. The properties of the upper sheet and fluid layer are  bending resistance $s_1=0.8 Pa m^3$, sheet thickness $b_1=1 cm$, sheet density $\rho_1=954 kg/m^3$, liquid viscosity $\mu=60 Pa\cdot s$,  liquid gap height $2h_0=5 mm$, and liquid density $\rho_l=750 kg/m^2$. In panel (a) the bottom surface is rigid. In panel (b) the bottom surface is an elastic spring array with $s_k=12 GPa/m$. In panel (c) the bottom surface is an elastic spring array with $s_k=12 GPa/m$ and mass-per-area of $i_2=25Kg/m^2$. Red, green, and orange vertical lines correspond to resonance frequencies of the upper sheet, the lower substrate, and a combine reference configuration with a constraint of constant gap between the upper sheet and the lower surface.}
\label{fig_filter}
\end{figure}

Figure \ref{fig_filter} presents the magnitude and phase of  liquid pressure vs. excitation frequency $\omega$ for wavelengths of $l=0.06$, $0.08$ and $0.1m$, denoted by the solid, dashed, and dotted lines, respectively. Vertical lines denote resonance frequency of the reference solid configurations of the upper sheet (red), bottom surface (green) and two sheets with a constraint of constant gap (orange). In all panels (a-c) the properties of the upper sheet and fluid layer are defined by bending resistance $s_1=0.8 Pa m^3$, sheet thickness $b_1=1 cm$, sheet density $\rho_1=954 kg/m^3$, liquid viscosity $\mu=60 Pa\cdot s$,  liquid gap height $2h_0=5 mm$, and liquid density $\rho_l=750 kg/m^2$ (characteristic to rubber and silicon oil). In panel (a) the bottom surface is rigid. In panel (b) the bottom surface is an elastic spring array with $s_k=12 GPa/m$. In panel (c) the bottom surface is an elastic spring array with $s_k=12 GPa/m$ and mass-per-area of $i_2=25Kg/m^2$.

In all of the configurations, sufficiently small frequencies yield negligible liquid pressure amplitude. An intermediate range of frequencies, which include the resonance frequency of the elastic sheet, leads to liquid pressure with amplitude and phase identical to the external pressure. In configuration (a) large frequencies lead to decay of the liquid pressure due to growing dominance of the elastic sheet inertia. Thus, for $Z_2\rightarrow\pm\infty$, the response of the system is similar to a bandpass filter and the fluidic pressure cannot exceed the external excitation. In configuration (b), similar behaviour is observed with an additional peak of the fluidic pressure near the elastic resonance frequency defined by (\ref{Z_2_alpha_2_const})-(\ref{opt_max}) (at $\omega\approx 40KHz$). At greater frequencies both the liquid pressure and phase decay to zero. Configuration (c) involves an additional resonance frequency of the lower surface, yielding a minima of the pressure near the combined reference resonance $(Z_1+Z_2=0)$. 

\clearpage

\appendix
\section{Leading order equations for $d^{'*}/\bar d^*\ll1$}
For dynamics characterized by $d^{'*}/\bar d^*\sim\varepsilon_1^{2}\ll1$ order of magnitude yields that the convection terms $\rho(\boldsymbol u_\parallel,w) \cdot \left(\bnabla_\parallel,{\partial}/{\partial z}\right)(\boldsymbol u_\parallel,w)$ in equation (\ref{mom_eq}) scales as $R_e/\varepsilon_1$ and may not be neglected. In addition, the sheets mean acceleration $\partial \bar d^2/\partial t^2$ yields significant pressures gradients in the trasverse direction ($\partial p/\partial z$). Thus, the leading order governing equations (\ref{norm_mom_eq}) are of the form
\begin{subequations}
\label{mom_eq_largedbar}
\begin{equation}
\alpha^2\frac{\partial \boldsymbol U}{\partial T}+\frac{Re}{\varepsilon_1}\frac{\partial \bar D}{\partial T}\frac{\partial \boldsymbol U}{\partial Z}=-\bnabla P+\frac{\partial^2 \boldsymbol U}{\partial Z^2}+O\left(\varepsilon_1^2,\varepsilon_1Re,\right),\quad 
\end{equation}
\begin{equation}
\alpha^2\frac{\partial^2 \bar D}{\partial T^2}=-\frac{\partial P}{\partial Z}+O\left(\varepsilon_1^2,\varepsilon_1 Re\right),
\end{equation}
\begin{equation}
\frac{\partial W'}{\partial Z}+\bnabla\cdot\boldsymbol U=0,
\end{equation}
\end{subequations}
and leading order boundary conditions (\ref{norm_bc}) are
\begin{subequations}
\begin{equation}
W'(Z=H_0)+W'(Z=-H_0)=O(\varepsilon_2),
\end{equation}
\begin{equation}
W'(Z=H_0)-W'(Z=-H_0)=2\frac{\partial D'}{\partial T}+O(\varepsilon_2),
\end{equation}
\begin{equation}
\boldsymbol U(Z=H_0)= O(b_1k\varepsilon_1),\quad \boldsymbol U(Z=-H_0)\sim O(b_2k\varepsilon_1),
\end{equation}
\begin{equation}
\bigg[\left(-s_1k^4\nabla^4+t_1k^2 \nabla^2-i_1\omega^2\frac{\partial^2}{\partial T^2}\right)\frac{h_0}{\mu\omega}\bigg](\bar D+\varepsilon_1^2D')+P-P_e=0,
\end{equation}
\begin{equation}
\bigg[\left(-s_2k^4\nabla^4 +t_2k^2 \nabla^2  -s_k-i_2\omega^2\frac{\partial^2 }{\partial T^2} \right)\frac{h_0}{\mu\omega}\bigg](\bar D-\varepsilon_1^2D')-P =0.
\end{equation}
\label{bc_largedar}
\end{subequations}
Equations (\ref{mom_eq_largedbar})-(\ref{bc_largedar}) yields that in the leading order dynamics for $d^{'*}/\bar d^*\sim\varepsilon_1^{2}\ll1$, the liquid pressure is created by elastic displacements, but is not significant in determining displacement dynamics of steady-state solutions. 

\section{Results in dimensional form}
\label{app_dim_res}
We present here the steady-state oscillating solutions in dimensional form. The dimensional liquid pressure (\ref{phat_sol}) is

\begin{equation}
\label{p_dim_sol}
\hat p=\left(\begin{split} 
\frac{2h_0k^2(-s_1k^4-t_1k^2 +i_1\omega^2)}{\rho_l \omega^2}\left(1-\frac{\tanh(\sqrt{\rho_l h_0^2 \omega i/\mu})}{\sqrt{\rho_l h_0^2 \omega i/\mu}}\right)\\
+1 +\frac{-s_1k^4-t_1k^2 +i_1\omega^2}{-s_2k^4 -t_2k^2  -s_k + i_2\omega^2}
\end{split}\right)^{-1}\hat p_e,
\end{equation}
the dimensional longitudinal (\ref{uhat_sol}) and transverse (\ref{wtag_sol}) liquid velocities are
\begin{equation}
\hat u=\left(\frac{\cosh(\sqrt{\rho_l \omega i/\mu}z)}{\cosh(\sqrt{\rho_l \omega i/\mu}h_0)}-1\right)\frac{k\hat p}{\omega\rho_l}
\end{equation}
and
\begin{equation}
\hat w'=i\left(\frac{z}{h_0}-\frac{\sinh(\sqrt{\rho_l \omega i/\mu}z)}{\sqrt{\rho_l \omega i/\mu}h_0\cosh(\sqrt{\rho_l \omega i/\mu}h_0)}\right)\frac{ k^2 h_0 \hat p}{\omega \rho_l}.
\end{equation}
The dimensional relative deformation (\ref{dtag_sol}) is
\begin{equation}
\hat d'=\left(1-\frac{\tanh(\sqrt{\rho_l \omega i/\mu}h_0)}{\sqrt{\rho_l \omega i/\mu}h_0}\right)\frac{k^2 h_0\hat p}{\omega^2 \rho_l }
\end{equation}
and the dimensional mean sheet deformation (\ref{dbar_sol}) is
\begin{equation}
\bar d=\left(1-\frac{\tanh(\sqrt{\rho_l \omega i/\mu}h_0)}{\sqrt{\rho_l \omega i/\mu}h_0}+\frac{\mu\omega}{h_0^3k^2(-s_2k^4 -t_2k^2  -s_k + i_2\omega^2)}\right)\frac{h_0 k^2\hat p}{\rho_l \omega^2}
\label{dbar_dim_sol}
\end{equation}

The three-dimensional response dynamics for an arbitrary external pressure field is therefore given in dimensional form by \citep{arfken2011mathematical}
\begin{equation}
\left(\begin{matrix} 
p\\
u_x\\
v_y\\
w'\\
\bar d\\
d'\\
\end{matrix}\right)=\frac{1}{(2\pi)^{3/2}}\int_{-\infty}^{\infty}{\int_{-\infty}^{\infty}\int_{-\infty}^{\infty}\hat{\left(\begin{matrix} 
\hat p\\
k_x \hat u/k\\
k_y \hat u/k\\
\hat w'\\
\hat {\bar d}\\
\hat d '\\
\end{matrix}\right)e^{-i(k_xx+k_yy+\omega t)}\,\,\,}dk_x dk_y d\omega},
\label{all_four_trans}
\end{equation}
where $\sqrt{k_x^2+k_y^2}=k$, and $u_x,u_y$ denote the $x,y$ direction velocity components and $\hat p_e$ is calculated from the inverse transformation
\begin{equation}
\hat p_e(k_x,k_y,\omega)=\frac{1}{(2\pi)^{3/2}}\int_{-\infty}^{\infty}\int_{-\infty}^{\infty}\int_{-\infty}^{\infty}p_e(x,y,t)e^{i(k_xx+k_yy+\omega t)} dxdydt.
\label{pe_inverse}
\end{equation}
Equations (\ref{all_four_trans}) and (\ref{pe_inverse}) require the dimensional form since the normalization of the two-dimensional problem was dependent on $k$ and $\omega$.


\section{Extrema points of $\hat D'(Z_1,Z_2,\alpha^2)$, $\hat {\bar D}(Z_1,Z_2,\alpha^2)$}
We obtain extrema points of the magnitude of $\hat D'=F \hat P$. The solution 
consists of the liquid pressure multiplied by $F=F_1+iF_2=(\alpha^2)^{-1}(H_0-\tanh(\sqrt{\alpha^2iH_0})/\sqrt{\alpha^2i})$. Thus, extrema points with respect to $Z_1$, $Z_2$ (where the remaining variables are kept constant) will yield the same expressions as for the liquid pressure - Eq. (\ref{Z_2_alpha_2_const}) and (\ref{z1_max_diff_z2}), respectively. We now obtain extrema points with respect to $\alpha^2$. We simplify the problem by investigating
\begin{equation}
\bigg|\frac{\hat P_e}{\hat D'}\bigg|^2=\frac{(2Z_1F_1+1+Z_1Z_2^{-1})^2+(2Z_1F_2)^2}{F_1^2+F_2^2}.
\end{equation}
Differentiating with respect to $\alpha^2$ and equating to zero yields
\begin{equation}
Z_1\left(4\frac{\partial F_1}{\partial \alpha^2}|F|^2-(4F_1+Z_2^{-1})\frac{\partial |F|^2}{\partial \alpha^2}\right)-\frac{\partial |F|^2}{\partial \alpha^2}=0.
\label{dtag_ex_alph}
\end{equation}
We now obtain extrema points of the magnitude of $\hat {\bar D}=(F+Z_2^{-1})\hat P$. The equation defining the extremum point of $\hat {\bar D}$ with respect to $Z_1$ is the same as the liquid pressure and relative deformation and is defined by (\ref{Z_2_alpha_2_const}). Next we obtain an expression for the extrema point of $\hat {\bar D}$ for $\alpha^2$. We simplify the problem by investigating
\begin{equation}
\bigg|\frac{\hat P_e}{\hat {\bar D}}\bigg|^2=\frac{(2Z_1F_1+1+Z_1Z_2^{-1})^2+(2Z_1F_2)^2}{(F_1+Z_2^{-1})^2+F_2^2}.
\label{dbartope_magsq}
\end{equation}
By differentiating \ref{dbartope_magsq} with respect to $\alpha^2$ and equating to zero yields an inexplicit relationship
\begin{equation}
A Z_1^2+BZ_1+C=0,
\label{quad_z1}
\end{equation}
where $A=\partial_{\alpha^2} ((2F_1+Z_2^{-1})^2+(2F_2)^2)\times((F_1+Z_2^{-1})^2+F_2^2)^{-1}$, $B=\partial_{\alpha^2} (4F_1+Z_2^{-1})\times((F_1+Z_2^{-1})^2+F_2^2)^{-1}$, and $C=\partial_{\alpha^2} ((F_1+Z_2^{-1})^2+F_2^2)^{-1}$. We may solve (\ref{quad_z1}) as a quadratic equation in terms of $Z_1$ and obtain a functional relationship between $Z_1$ to $Z_2$ and $\alpha^2$.

\section{Comparison between dynamics of two sheets connected by a stiff spring array to an elastic Hele-Shaw cell}
We here calculate the phase and amplitude of the steady state oscillations for a reference configuration consisting of two elastic sheets connected by a spring array. The response of two elastic sheets with a constraint of constant gap or an isolated upper sheet is obtained directly from limits of the spring array coefficient.    

The governing equation of the upper elastic sheet is
\begin{equation}
-s_1\frac{\partial^4 d_1}{\partial x^4}+t_1\frac{\partial^2 d_1}{\partial x^2}-s_{12}(d_1-d_2)-p_e=i_1\frac{\partial^2 d_1}{\partial t^2},
\end{equation}
and the governing equation of the lower elastic sheet is 
\begin{equation}
-s_2\frac{\partial^4 d_2}{\partial x^4}+t_2\frac{\partial^2 d_2}{\partial x^2}-s_k d_2+s_{12}(d_1-d_2)=i_2\frac{\partial^2 d_2}{\partial t^2},
\end{equation}
where $s_{12}$ is the spring stiffness connecting the sheets. We substitute the wave form
\begin{equation}
f=\hat f e^{i(k x+\omega t)},\quad f=d_1, d_2, p_e,
\end{equation}
 for all variables, and obtain the upper sheet deformation 
\begin{equation}
\hat d_1=\frac{\tilde z_2-s_{12}}{(\tilde z_2-s_{12})(\tilde z_1-s_{12})-s_{12}^2}\hat p_e,
\label{spconfg_d1}
\end{equation}
and the lower sheet deformation
\begin{equation}
\hat d_2=\frac{-s_{12}}{(\tilde z_2-s_{12})(\tilde z_1-s_{12})-s_{12}^2}\hat p_e.
\label{spconfg_d2}
\end{equation}
where $\tilde z_n=-s_nk^4-t_nk^2+i_n\omega^2-s_k(n-1),\quad n=1,2$.
From Eq. (\ref{spconfg_d2}), and (\ref{spconfg_d1}) we see that resonance is obtained when
\begin{equation}
(\tilde z_2-s_{12})(\tilde z_1-s_{12})-s_{12}^2=0
\end{equation}
isolating $\tilde z_1$ we obtain
\begin{equation}
\tilde z_1=\left(\frac{1}{s_{12}}-\frac{1}{\tilde z_2}\right)^{-1}
\end{equation}
for the case the spring is much stiffer than the sheets we obtain that resonance will occure at $\tilde z_1=-\tilde z_2$

\acknowledgments{This research was supported by the ISRAEL SCIENCE FOUNDATION (Grant No. 818/13).}

\bibliographystyle{jfm}
\bibliography{main_2.bbl}

\end{document}